\documentclass[lettersize,journal]{IEEEtran}
\usepackage{amsmath}
\usepackage{amssymb}
\usepackage{amsfonts}

\newcommand{\Rmnum}[1]{\expandafter\@slowromancap\romannumeral #1@}
\usepackage{graphicx}
\usepackage{epsfig}
\usepackage{subfigure}
\usepackage{graphicx}

\usepackage{citesort}
\usepackage{xcolor}
\usepackage{enumerate}
\usepackage{extarrows}
\usepackage{algorithmic}
\usepackage{tensor}
\usepackage{booktabs}
\usepackage{float}  

\begin{document}
	\title{Unsupervised Learning for AoD Estimation in MISO Downlink LoS Transmissions}
	
\author{Jiaying Li,  Yuanwei Liu, and Hong Xing
	\thanks{J. Li and H. Xing are with the IoT Thrust, The Hong Kong University of Science and Technology (Guangzhou), Guangzhou, 511453, China; H. Xing is also affiliated with the Department of ECE, The Hong Kong University of Science and Technology, HK SAR. (e-mails: jli989@connect.hkust-gz.edu.cn,~hongxing@ust.hk).
		Y. Liu is with the Department of EEE, The
		University of Hong Kong, HK SAR.  (e-mail: yuanwei@hku.hk).}
}
\maketitle

\begin{abstract}
	 With the emergence of simultaneous localization and communication (SLAC), it becomes more and more attractive to perform angle of departure (AoD) estimation at the receiving Internet of Thing (IoT) user end for improved positioning accuracy, flexibility and enhanced user privacy. To address challenges like a large number of real-time measurements required for latency-critical applications and enormous data collection for training deep learning models in conventional AoD estimation methods, we propose in this letter an unsupervised learning framework, which unifies training for both deterministic maximum likelihood (DML) and stochastic maximum likelihood (SML) based AoD estimation in multiple-input single-output (MISO) downlink (DL) wireless transmissions.
			Specifically, under the line-of-sight (LoS) assumption, we incorporate  both the received signals and pilot-sequence information, as per its availability at the DL user, into the input of the deep learning model, and adopt a common neural network architecture compatible with input data in both DML and SML cases. Extensive numerical results validate that the proposed unsupervised learning based AoD estimation not only improves estimation accuracy, but also significantly reduces required number of observations, thereby reducing both estimation overhead and latency compared to various benchmarks.
\end{abstract}

\IEEEpeerreviewmaketitle
\newtheorem{definition}{\underline{Definition}}[section]
\newtheorem{fact}{Fact}
\newtheorem{assumption}{Assumption}
\newtheorem{theorem}{\underline{Theorem}}[section]
\newtheorem{lemma}{\underline{Lemma}}[section]
\newtheorem{proposition}{\underline{Proposition}}[section]
\newtheorem{corollary}[proposition]{\underline{Corollary}}
\newtheorem{example}{\underline{Example}}[section]
\newtheorem{remark}{\underline{Remark}}[section]
\newtheorem{algorithm}{\underline{Algorithm}}[section]
\newcommand{\mv}[1]{\mbox{\boldmath{$ #1 $}}}
\newcommand{\mb}[1]{\mathbb{#1}}
\newcommand{\Myfrac}[2]{\ensuremath{#1\mathord{\left/\right.\kern-\nulldelimiterspace}#2}}
\newcommand\Perms[2]{\tensor[^{#2}]P{_{#1}}}


\section{Introduction}
\IEEEPARstart{S}{imultaneous} localization and communication (SLAC), which integrates localization and communication using communication infrastructure for localization, has recently received significant attention due to its higher refresh rate, improved coverage, and better positioning accuracy compared to the Global Positioning System (GPS) \cite{gong2023high}. In SLAC, leveraging communication purposed signaling, key location-related parameters such as time of arrival (ToA), time difference of arrival (TDoA), angle of arrival (AoA) and Doppler shift can be extracted from receiving measurements. Among a variety of techniques to implement SLAC, in the past, positioning with uplink (UL) transmissions was a common practice as it directly extracts position related parameters from channel estimation \cite{gong2023high}. By contrast, SLAC based on downlink (DL) transmissions, supported by 3GPP Release 16 and 17 \cite{wymeersch2021integration}, becomes prevalent  thanks to improved energy efficiency and signal-to-noise ratio (SNR) for  parameter estimation performed at mobile devices. Furthermore, DL transmission based SLAC provides flexibility for self-positioning at mobile devices, allowing for positioning using different types of signals (e.g., pilots or information) from multiple base stations or anchors, and is also conducive to preservation of users' privacy.

Despite these benefits, most Internet of Things (IoT) devices are equipped with single antenna and limited communication capacities, thus making conventional subspace-based angle estimation methods leveraging spatial high resolution, such as MUSIC \cite{1986MUSIC}, ESPRIT \cite{roy1989esprit}, as well as Capon \cite{capon1969high}, which achieves minimum variance unbiased estimation of the transmit signal, not viable anymore. In literature, existing angle estimation methods that do not rely on multiple receiving antennas encompass beam scanning based on received signal strength (RSS) \cite{giorgetti2009single}, maximum likelihood (ML) estimation \cite{jaffer1988maximum} and channel estimation based methods \cite{ma2022channel}. For example, in \cite{ma2022channel}, the authors proposed first estimating the channel using the least squares (LS) estimator and then employing the MUSIC algorithm to estimate the angle from the estimated channel.
As for angle of departure (AoD) estimation in multiple-input single-output (MISO) DL line-of-sight (LoS) transmission, ML estimators are often employed \cite{fascista2019millimeter,wang2021joint,keykhosravi2022ris} for improved estimation accuracy and better noise robustness. For instance, the authors in \cite{fascista2019millimeter} used grid search with an ML estimator for AoD estimation in mmWave wireless systems. Reconfigurable intelligence surface (RIS)-assisted indoor scenarios with blocked LoS paths were considered in \cite{wang2021joint}, where a similar grid search approach was adopted to estimate AoD from the RIS to the user, followed by gradient descent refinement to improve accuracy. Reference \cite{keykhosravi2022ris} performed offline discrete Fourier transform (DFT) of pilot signals, which transforms the original estimation problem into searching for the optimal DFT index, and then fine tuning the estimated AoD leveraging the quasi-Newton algorithm.

On another front, rapid advancement of deep learning witnesses the paradigm shift for angle estimation using data-driven methods with improved estimation accuracy, real-time inference capability, and possibly relaxed system assumptions, which are particularly attractive to complex wireless transmissions environments with low SNR or high mobility \cite{liu2018direction,yang2021machine}. Angle estimation methods based on \emph{supervised learning} use a large amount of labeled training data, e.g., true angles, to train classification \cite{liu2018direction} or regression \cite{yang2021machine} models that learn the mapping between input signals and target angles. To alleviate the cost of enormous data collection with labeling, \emph{unsupervised learning} approaches have recently been explored to solve angle estimation problems \cite{yuan2021unsupervised,weisser2023unsupervised}.  In \cite{yuan2021unsupervised}, the authors applied unsupervised learning to AoA estimation using a loss function that captures the squared Frobenius-norm difference between the received signal's sample covariance and the reconstructed covariance matrix. However, this approach leads to suboptimal AoA estimation accuracy due to neglect of the underlying statistical properties of the transmit signal, resulting in a possible mismatch between the employed loss function and the objective of AoA angle estimation, while reference \cite{weisser2023unsupervised}  proposed unsupervised learning with ML directly serving as the training objective, thereby improving the AoA estimation accuracy.

As mentioned above, both model-based and data-driven methods have, nevertheless, their limitations in AoD estimation for MISO DL transmissions. On one hand, conventional model-based methods \cite{fascista2019millimeter,wang2021joint,keykhosravi2022ris} require a large number of measurements, which is not suitable for latency-critical applications. On the other hand, existing unsupervised learning approaches, e.g.,  \cite{weisser2023unsupervised}, considered using only sample covariance as the input of the neural network and the stochastic maximum likelihood (SML) case, the angle-estimation accuracy of which can be significantly improved when the pilot-sequence information is included. This motivates us to propose an unsupervised learning framework that unifies training for both deterministic ML (DML) and SML based AoD estimation in MISO DL transmissions.  Our contributions are summarized as follows. 1)~We incorporate both the received signals and pilot-sequence information, as per its availability at the DL user, into the input of the deep learning model. 2)~We adopt a common neural network architecture compatible with training data for both DML and SML based AoD estimation.
3)~We show by numerical results that the proposed unsupervised learning based AoD estimation not only improves estimation accuracy, but also requires significantly fewer observations when being deployed for inference, thereby reducing both estimation overhead and latency compared to existing benchmarks.

The remainder of this paper is organized as follows. The system model and problem formulation are presented in Section \ref{sec:System Model and  Problem formulation}. Section \ref{section3} introduces our proposed unified unsupervised learning framework for ML based AoD estimation. The simulation results are provided in Section \ref{exp} followed by conclusions and remarks drawn in Section \ref{conclusion}.

\begin{figure*}[h]
	\centering
	\includegraphics[width=.9\textwidth]{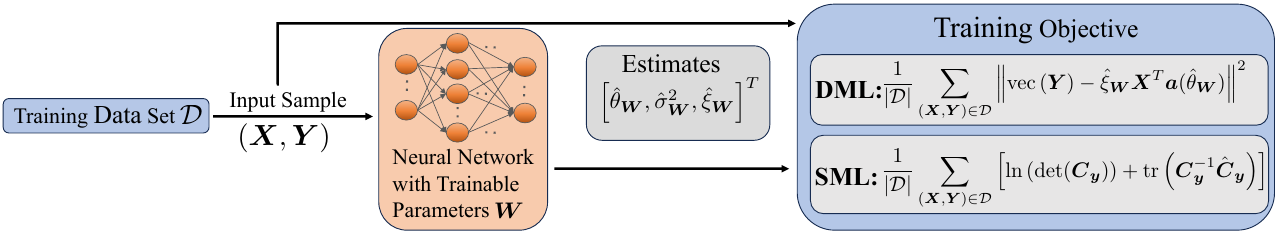}
	\caption{Illustration of the proposed unsupervised learning framework.}
	\label{fig:DL_model}
\end{figure*}

\section{System Model and Problem Formulation}\label{sec:System Model and  Problem formulation}
\subsection{System Model}
We consider DL narrow-band wireless communications via LoS transmissions from a base station (BS), equipped with a uniform linear array (ULA) of $M$ antennas, to a user equipment (UE), equipped with an omni-direction antenna. The AoD from the BS to the UE is denoted as $\theta$.  Under the assumption of narrow-band and far-field transmission, the steering vector $\mv{a} (\theta) \in \mathbb{C}^{M\times 1}$ departing from the BS is given by 
\begin{align}
	\mv{a} (\theta) = \begin{bmatrix}
		1, e^{-j\frac{2\pi d}{\lambda}\cos{\theta}}, \ldots, e^{-j\frac{2\pi(M-1)d}{\lambda}\cos{\theta}}
	\end{bmatrix}^T, \label{steering}
\end{align}
where $\lambda$ denotes the wavelength, and $d$ represents the spacing between adjacent elements of the ULA.  

We divide the total pilot transmission duration $T$ into $Q$ equal-length blocks, each of which is further divided into $L$ time slots of unit length for pilot transmissions, a.k.a., $T=QL$. 
 Denote the $\mv{X}^{(q)} \triangleq [
\mv{x}_1^{(q)},...,\mv{x}_L^{(q)} 
]^T \in \mathbb{C}^{L \times M}$ as the pilot sequence transmitted at block $q$,   $q \in \left\{1,...,Q\right\} \triangleq [Q]$. The transmit pilot $\mv{x}_l^{(q)}$ at the $l$-th time slot, $l\in \left\{ 1,\ldots,L\right\}\triangleq[L]$, is, without loss of generality (w.l.o.g), expressed as $\mv{x}_l^{(q)} =c^{(q)} \tilde{\mv x}_l $, where \(c^{(q)}\sim \mathcal{CN}(0,1)\) is  the transmit signal of block $q$, and $\tilde{\mv x}_l$ is a given fixed beamforming vector such that $\|\tilde{\mv x}_l\|^2\le P$. It is worth noting that multiple time slots ($L>1$) are required to create the covariance matrix of the received signal $\mv y^{(q)}$,  instead of a variance (scalar), thus facilitating estimation of the $\theta$ in the sequel.
Assuming that the UE's position remains static over the $Q$ blocks, the received signal in block $q$, \(q\in [Q]\), can be expressed as
\begin{align}
	\mv{y}^{(q)} \triangleq \begin{bmatrix}
		{y}^{(q)}_1,\ldots 	{y}^{(q)}_L
	\end{bmatrix}^T =  \xi \mv{X}^{(q)}  \mv{a}\left(\theta \right) + \mv{w}^{(q)}, \label{receive_signal}
\end{align}
where {$\mv{w}^{(q)}$ is the additive white Gaussian noise (AWGN), denoted by $\mv{w}^{(q)} \sim \mathcal{CN}\left(\mv{0},\sigma^2 \mv{I}\right)$}, and \textcolor{black}{$\xi$ is the channel gain parameter determined by factors such as random phase shifts and the range between the BS and the UE, etc.}

\subsection{Maximum Likelihood (ML) based Problem Formulation}
As for estimations of the AoD $\theta$ and the parameter $\xi$ at the UE, we adopt the ML estimation, which can be divided into two types based on the pilot information known at the UE \cite{Summary_DOA}:
\subsubsection{Deterministic ML (DML)} If the UE fully knows the pilot $\mv{X}^{(q)}$ sequences over all blocks, $q \in [Q]$, the received signal $\mv{y}^{(q)}$ in \eqref{receive_signal} follows complex Gaussian distribution, denoted by $\mv{y}^{(q)}\sim \mathcal{CN} ( \xi \mv{X}^{(q)} \mv{a}(\theta), \sigma^2 \mathbf{I} )$. Since $\mv{w}^{(q)}$ is independent and identically distributed ($i.i.d.$) across all blocks, it can be easily shown that the ML estimation for $\theta$ and $\xi$ for maximizing the log-likelihood $p(\mv{y}^{(1)},...,\mv{y}^{(Q)};\theta,\xi,\sigma^2)$ is equivalent to solving a LS problem as follows \cite{kay1993fundamentals}:
\begin{align}
	\mathop{\mathtt{Minimize}}_{\theta,~\xi}~~   \sum \limits_{q=1}^Q \left\| \mv{y}^{(q)} - \xi   \mv{X}^{(q)}  \mv{a}\left(\theta \right)   \right\|^2_2 \label{problem_DML}.
\end{align}
\subsubsection{Stochastic ML (SML)}
In contrast with DML, SML assumes that the UE only knows partial information about the pilot sequences $ \{\mv x_l^{(q)}\}_{l=1}^L$, for all $q\in[Q]$. Specifically, the UE knows the first order statistics $\mathbb{E} [\mv{x}_l^{(q)}]=\mv{0}$, and the second order statistics $  \boldsymbol{R}_{l l^{\prime}}=\mathbb{E}[\mv{x}_l^{(q)}(\mv{x}_{l^{\prime}}^{(q)})^H] = \tilde{\boldsymbol{x}}_l \tilde{\boldsymbol{x}}_{l^{\prime}}^H $,  $l, l^\prime \in [L]$. 
	As a result, the received signal $\mv{y}^{(q)}$ follows a complex Gaussian distribution, i.e., $\mv{y}^{(q)} \sim \mathcal{CN} (\mv{0}, \mv{C}_y)$, where $\mv{C}_y$ denotes the covariance matrix, given by
\begin{align}
\notag	\mv{C}_y &=\left|\xi\right|^2 \mathbb{E} \left[   \mv{X}^{(q)}  \mv{a}\left(\theta \right)\mv{a}^H\left(\theta \right)\left( \mv{X}^{(q)}\right)^H  \right] + \sigma^2 \mv{I}  \\&= \mv{R}_{\boldsymbol{X}}(\theta)+ \sigma^2 \mv{I},  \label{Cy_cal}
\end{align}
in which each element of $\mv{R}_{\boldsymbol{X}}(\theta) \in \mathbb{C}^{L \times L}$ can be expressed as, $l, l^\prime\in[L]$,
\begin{align}
	\left[\mv{R}_{\boldsymbol{X}}(\theta) \right]_{l,l^\prime} = \left|\xi\right|^2 \mv{a}^{T}\left(\theta\right) \mv{R}_{ll^\prime}\mv{a}^{*}\left(\theta\right). \label{R_x}
\end{align}
Similar to the DML case, under the assumption of $i.i.d.$ pilot transmission over $q\in [Q]$, the ML estimation problem in terms of the log-likelihood $p(\mv y^{(1)},\ldots, \mv y^{(Q)}; \theta, \xi,\sigma^2)$ is equivalent to the minimization problem \cite{weisser2023unsupervised}:
\begin{align}
	\mathop{\mathtt{Minimize}}_{\theta,~\xi,~\sigma^2}~~ \left[\ln \left(\operatorname{det}\left(\boldsymbol{C}_{{y}}\right)\right)+\operatorname{tr}\left(\boldsymbol{C}_{{y}}^{-1} \hat{\boldsymbol{C}}_{{y}}\right)\right], \label{problem_SML}
\end{align}
where $\hat{\boldsymbol{C}}_{\boldsymbol{y}}$ denotes the sample covariance matrix over $Q$ blocks, i.e.,
\begin{align}
	\hat{\boldsymbol{C}}_{\boldsymbol{y}}=\frac{1}{Q} \sum_{q=1}^Q \boldsymbol{y}^{(q)} \left(\boldsymbol{y}^{(q)}\right)^H.
	\label{sample_C_y}
\end{align}

Solving the formulated ML estimation in problems \eqref{problem_DML} and \eqref{problem_SML} for \(\theta\) and \(\xi\) typically involves a two dimension search~\cite{kay1993fundamentals}, which incurs high computational costs, thus detracting from real-time application capabilities. For the DML case, the prevailing method employs DFT of the pilot sequence $\mv{x}_l^{(q)}$ with the number of DFT points being greater than the number of antennas, which simplifies the two dimension search to one dimension \cite{keykhosravi2022ris}. This method nevertheless still requires a sufficient number of observations, making latency-critical applications prohibitive. Recent work in \cite{weisser2023unsupervised} using auto-encoder (AE) based architectures with ML loss for training has enabled unsupervised learning to estimate angles, reducing the need for extensive observations. However, \cite{weisser2023unsupervised} only considered SML for uplink transmissions.

%

\section{A Unified Unsupervised Learning Framework for ML based AoD Estimation}\label{section3}
In this section, we propose an unsupervised learning framework for AoD estimation in LoS DL transmissions  at single-antenna UEs, unifying both cases as shown in
Fig.~\ref{fig:DL_model}, which consists of a training data set denoted by $\mathcal{D}$, a deep learning neural network (NN), e.g., convolutional NN (CNN), in which a common network architecture is shared by both SML and DML, and a training objective customized to these two different types of ML estimation. Unlike \cite{weisser2023unsupervised}, where the input feature is just the sample covariance \(\hat{\boldsymbol{C}}_y\) defined in \eqref{sample_C_y}, our proposed framework takes both $\mv{X}$ and \(\boldsymbol{Y} \) as input samples, where \(\boldsymbol{Y}\triangleq [ \boldsymbol{y}^{(1)}, \ldots, \boldsymbol{y}^{(Q)}] \) represents the received observations over \(Q\) blocks, and \(\mv{X}\) is generated based on the information known about the pilot sequences 
at the UE, which is given by 
\begin{align} 
	\mv X =\left\{ \begin{array}{ll}
\begin{bmatrix}
	\left(\mv{X}^{(1)}\right)^T,\ldots, \left(\mv{X}^{(Q)}\right)^T
\end{bmatrix}		,\; &\mbox{DML case}\\
	\mv{1}^T \otimes  \begin{bmatrix}
			\mv{\tilde{x}}_1,...,\mv{\tilde{x}}_L 
	\end{bmatrix}
		,\; &\mbox{SML case},
	\end{array}\right. \label{pilot_case}
	\end{align}
 where $\mathbf{1}$ is a $Q$-dimension all-one vector, and $\otimes$ denotes the Kronecker product. Then, we reshape each data sample $(\mv{X}, \boldsymbol{Y}) \in\mathcal{D}$ into a tensor, given by: 
\begin{align}
\notag	(\mv{X}, \boldsymbol{Y}) \xrightarrow{\text {reshape}} \left[\begin{array}{ccccc}
		Y_{1,1} & \cdots & Y_{L, 1} & \cdots & Y_{L, Q} \\
		X_{1,1} & \cdots & X_{ 1,L} & \cdots & X_{1,QL} \\
		X_{2,1} & \cdots & X_{2,L} & \cdots & X_{2,QL} \\
		\vdots & \cdots & \vdots & \cdots & \vdots \\
		X_{M,1} & \cdots & X_{M,L} & \cdots & X_{M,QL}
	\end{array}\right] \triangleq \boldsymbol{S}, 
\end{align}
where $Y_{l, q}$ is the $(l, q)$-th entry of $\boldsymbol{Y}$ and $X_{m,l}$ is the $(m,l)$-th entry of $\mv{X}$, respectively. The tensor $ \boldsymbol{S}$ is then split into real and imaginary parts to form the input of the neural network. Next, the neural network operates as a function $\boldsymbol{f}(\cdot; \boldsymbol{W})$ with the trainable parameters denoted by $\boldsymbol{W}\in\mathbb{R}^{d\times 1}$, mapping from the input data sample $(\mv{X}, \mv{Y})$ to the estimates of the AoD, the noise variance, and the channel gain, denoted by
$ \hat{\theta}_{\boldsymbol{W}}, \hat{\sigma}_{\boldsymbol{W}}^2$ and $\hat{\xi}_{\boldsymbol{W}}$, respectively, which can thus be expressed as
$\boldsymbol{f} (\mv{X}, \boldsymbol{Y}; \boldsymbol{W}) = [ \hat{\theta}_{\boldsymbol{W}}, \hat{\sigma}_{\boldsymbol{W}}^2, \hat{\xi}_{\boldsymbol{W}}]^T$.

Next, we elaborate on training data sets generation and training objectives for problem formulation in cases of DML and SML, respectively. 
\\ \noindent \emph{1) DML: }
If the UE fully knows the pilot sequences over all $Q$ blocks, i.e., {$\{\mv{X}^{(q)}\}_{q=1}^Q$}, the UE can reconstruct the noiseless received signal from the output of the neural network as $\hat{\xi}_{\boldsymbol{W}} \mv{X}^{(q)} \mv{a}(\hat{\theta}_{\boldsymbol{W}})$, $q\in [Q]$. The empirical training loss minimization in accordance with problem \eqref{problem_DML} is thus given by:
\begin{align}
	\mathop{\mathtt{Minimize}}_{\boldsymbol{W}}~~ \frac{1}{|\mathcal{D}|}\sum \limits_{\left(\boldsymbol{X},\boldsymbol{Y}\right) \in \mathcal{D}} \left\| \operatorname{vec}\left(\mv{Y}\right)-\hat{\xi}_{\boldsymbol{W}} \mv{X}^T \mv{a}(\hat{\theta}_{\boldsymbol{W}})   \right\|^2. \label{loss_DML}
\end{align}
where $\operatorname{vec}(\boldsymbol{Y})$ and $|\mathcal{D}|$ denote the vectorization of matrix $\boldsymbol{Y}$ and the size of data set $\mathcal{D}$, respectively. \\
\noindent \emph{2) SML: }
If the UE only knows partial information about the pilots, i.e., {the given beamforming vector $\mv{\tilde{x}}_{l}$, \(l \in[L]\), the UE reconstructs the covariance matrix $\mv{C}_{\boldsymbol{y}}$ using \eqref{Cy_cal} and \eqref{R_x} via the estimates of the neural network $\hat{\theta}_{\boldsymbol{W}}, \hat{\sigma}_{\boldsymbol{W}}^2$ and $\hat{\xi}_{\boldsymbol{W}}$, and the sample covariance $\hat{\boldsymbol{C}}_{\boldsymbol{y}}$ using \eqref{sample_C_y} via $\mv{Y}$. In line with problem \eqref{problem_SML}, the empirical training loss minimization in the SML case is formulated as follows:
	\begin{align}
		\mathop{\mathtt{Minimize}}_{\boldsymbol{W}}~~ \frac{1}{|\mathcal{D}|} \sum_{\left(\boldsymbol{X},\boldsymbol{Y}\right) \in \mathcal{D}} \left[\ln \left(\operatorname{det}(\boldsymbol{C}_{\boldsymbol{y}})\right) + \operatorname{tr}\left(  \boldsymbol{C}_{\boldsymbol{y}}^{-1} \hat{\boldsymbol{C}}_{\boldsymbol{y}} \right)\right]. \label{loss_SML}
	\end{align}
	
The model parameters $\mv W$ in the proposed learning framework can then be trained by, e.g., stochastic gradient descent (SGD), to solve problems for DML (c.f.~\eqref{loss_DML}) and SML (c.f.~\eqref{loss_SML}), respectively.

It is noteworthy that the DML and the SML cases are unified in the proposed learning framework in the following senses. First, in terms of assumption about the knowledge of pilot sequence, DML can be considered as a special case of the SML, the UE in the latter of which can only access the first and second order moments of the pilot symbol $\mv{x}^{(q)}_l$. Secondly, in terms of the problem formulation, they aim at solving the same ML estimation problem, i.e., $\max_\theta p(\mv y^{(q)}; \theta,\xi,\sigma^2)$. In addition, they share the same dimension of features (c.f.~\eqref{pilot_case}) as input of a common NN architecture.

\section{Numerical Results}\label{exp}
	In this section, we perform numerical experiments to verify the effectiveness of our proposed unsupervised learning framework for MISO LoS downlink transmissions.
	
The steering vector $\boldsymbol{a}(\theta)$ is generated using the number of antennas $M=8$, the AoD $\theta$ being drawn uniformly from $(0, \pi/2)$, denoted by $\theta \sim \mathcal{U}(0, \pi/2)$, and the antenna spacing $d$ is set to be $\lambda/2$ with carrier frequency $f_c=28$\,GHz; the channel gain parameter $\xi$ is given by $\xi =\sqrt{\left({c}/{4 \pi f_c  r_0}\right)^2 \cdot \left({r_0}/{r}\right)^\gamma}$ \cite{goldsmith2005wireless} with the range $r \sim \mathcal{U}(20,50)$\,m, the reference distance $r_0 = 1$\,m, and the path-loss exponent $\gamma = 3$; the AWGN variance $\sigma^2$ set to be $-165$ dBm/Hz over a 120\,KHz bandwidth. {The beamforming vector
	 $\tilde{\mv{x}}_l$ is generated via $\tilde{\mv{x}}_l = \sqrt{{P}/{M}} [e^{j \pi \varrho^{(l)}_1}, e^{j \pi \varrho^{(l)}_2}, \ldots, e^{j \pi \varrho^{(l)}_{M}}]^T$ with fixed $\varrho_1^{(l)}, \ldots, \varrho_M^{(l)}$ varying from slot to slot and remaining the same across blocks, $l\in [L]$, $q \in [Q]$}\footnote{Any other specific form of beamforming design is applicable.}.}{ Next, we sample $200$ AoDs $\theta \sim \mathcal{U}(0, \pi / 2)$ and $220$ ranges $r \sim \mathcal{U}(20,50)$\,m, and generate $5$ distinct sets of beamforming vectors $\{\tilde{\boldsymbol{x}}_l\}_{l=1}^{L}$ and $5$ distinct sets of transmit signals $\{c^{(q)}\}_{q=1}^{Q}$, where each $c^{(q)} \sim \mathcal{C N}(0,1)$, $q \in [Q]$.} By enumerating all combinations of $\theta, r$, $\{\tilde{\boldsymbol{x}}_l\}_{l=1}^L$ and $\{c^{(q)}\}_{q=1}^Q$ as well as  independent realizations of $\mv w^{(q)}\sim \mathcal{CN} (\mv{0},\sigma^2\mv{I})$, we generate $\boldsymbol{Y}$ as per  \eqref{receive_signal} and thus a total of $|\mathcal{D}|=1.1 \times 10^6$ data samples $(\boldsymbol{X}, \boldsymbol{Y})$ with $10^6$  for training, and $10^5$ for testing, respectively.
We employ {"\emph{ShuffleNetv2\_x\_05}"\footnote{{We choose "\emph{ShuffleNetv2\_x\_05}", among many others, for  its exceptional trade-off between predictive capability and computational cost, thus being a promising NN architecture for IoT UEs with limited computational capacity.}}} \cite{ma2018shufflenet} as the NN architecture described in Fig. \ref{fig:DL_model}, and the "\emph{AdamW}" optimizer \cite{loshchilov2017decoupled} with a mini-batch size of 2048 (without replacement) for training. In the test phase, the trained model is deployed at the UE to perform AoD estimation with the model output $\hat{\theta}_{\boldsymbol{W}}$ being evaluated by averaging over all data samples in the test data set. 
	
	We consider the following benchmarks: the \emph{DFT-based} method \cite{keykhosravi2022ris}, which performs a grid search to find the index with the smallest difference between the DFT of the fully known pilot $\mv{x}_l^{(q)}$ and the received signal $\mv y_l^{(q)}$, and is thus only applicable to the DML case; the \emph{AE-based} method \cite{weisser2023unsupervised} which also adopts unsupervised learning for AoD estimation but only considers the SML case. Since AoA estimation for narrow-band SIMO LoS uplink transmissions based on subspace techniques is well studied with provable performance guarantees \cite{analyse_music,li2022stability}, to demonstrate the viability of the proposed AoD estimation in the MISO LoS downlink scenario, we include the classic \emph{MUSIC }\cite{1986MUSIC} and \emph{ESPRIT} \cite{roy1989esprit} algorithms as baselines, where the UE transmits pilot sequence \(\{x^{(q)}\}_{q=1}^Q\), and the BS is equipped with a ULA of an equivalent of $L$ number of antennas for fair comparison.
	
			\begin{figure}[htp]
		\centering
			\includegraphics[width=3in]{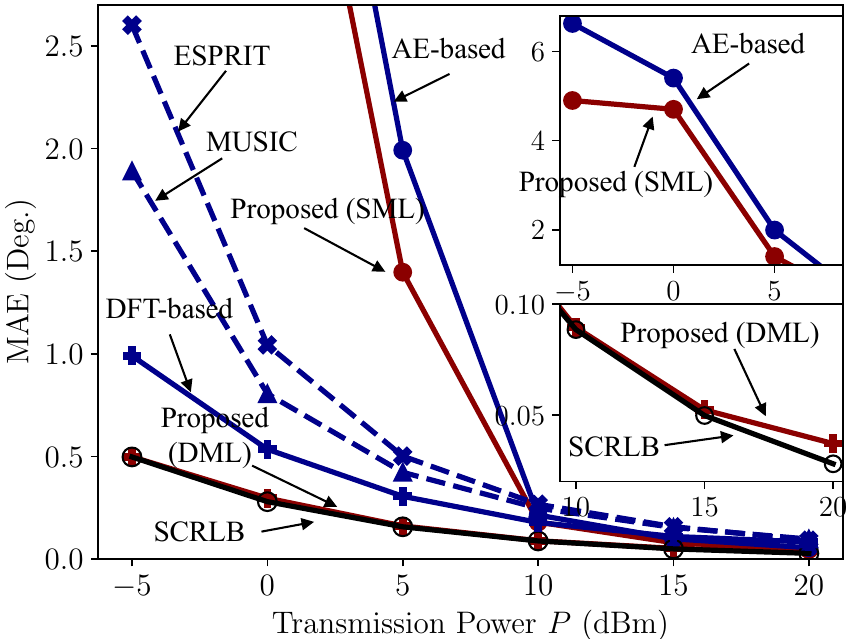}
		\caption{{The MAE versus  the transmit power $P$ with $L=6$, $Q=4$, $\theta=23.4^\circ$ and $r=32.1$\,m.}}
		\label{fig:changing_P}
	\end{figure}
	
First, we focus on estimation error by evaluating the mean absolute error (MAE) (in degree) between the estimated $\hat{\theta}_{\boldsymbol{W}}$ and the ground truth $\theta$, i.e., $\mathbb{E}[|\hat{\theta}_{\boldsymbol{W}} -\theta|]$, averaging over all test data set with fixed AoD $\theta$ and range $r$. Fig. \ref{fig:changing_P} shows that our proposed AoD estimation for the DML case not only 
	outperforms all other methods but also closely approaches the square root of the Cramér-Rao lower bound (SCRLB) with a negligible gap, and the schemes for the SML case perform less favorably, especially in low SNR regime, due to the unknown pilot symbols at the UE. \emph{MUSIC} and \emph{ESPRIT} methods render a diminishing gap  from the DML case with increased transmit power. Furthermore, when the transmit power is sufficiently large, all schemes approach the SCRLB with negligible gaps among each other, demonstrating the key role played by $P$ regardless of the level of information about pilots available at the UE. 
	
\begin{figure}[htp]
	\centering
	\includegraphics[width=3in]{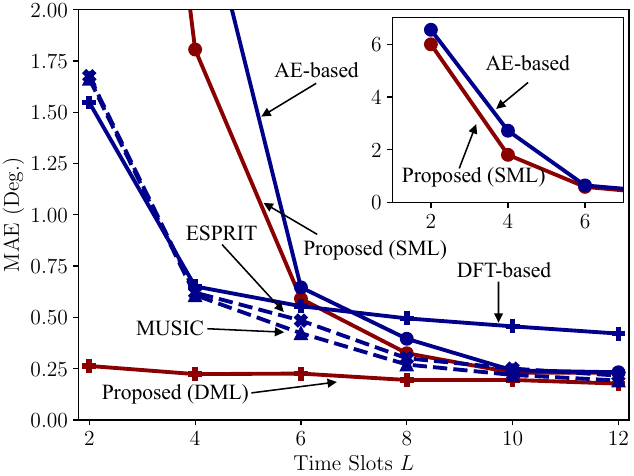}
	\caption{The MAE versus the number of the time slots $L$ in one block of pilot sequence with $P=15$\,dBm and $Q=4$.}
	\label{fig:changing_L}
\end{figure}

{Next,} Fig. \ref{fig:changing_L} demonstrates the impact of the number of
time slots $L$ on the MAE by different schemes. As expected, the proposed AoD estimation for the DML case outperforms all the other schemes, especially in conditions with fewer slots of transmissions in each block of the pilot sequence. Fig. \ref{fig:changing_L} shows that, in the DML case, as few as $T=QL=16$ observations is sufficient for accurate downlink AoD estimation at the UE, while a merely double number of observations achieves nearly the same performance as the equivalent uplink AoA estimation in the SML case. These results highlight the effectiveness of the proposed unified unsupervised learning framework in terms of significant pilots saving in DML cases, and improved trade-offs between computation complexity and estimation accuracy in SML cases.
	
 \begin{table}[h]
	\caption{Run time of estimation methods } 
	\label{tab:running_time} 
	\begin{tabular}{|p{.8cm}|p{.8cm}|p{.8cm}|p{.8cm}|p{.8cm}|p{.8cm}|p{.8cm}|}
		\hline
		Method           & Proposed (DML)         & DFT                    & Proposed (SML)         & AE-based               & ESPRIT                 & MUSIC                  \\ \hline
		Run Time& $9.557 \times 10^{-3}$\,s & $1.801 \times 10^{-3}$\,s & $9.161 \times 10^{-3}$\,s & $9.138 \times 10^{-3}$\,s & $8.671 \times 10^{-4}$\,s & $4.671 \times 10^{-2}$\,s \\ \hline
	\end{tabular}
\end{table}

	Finally, to evaluate the wall-clock running time of all above approaches, the numerical results are obtained using Python 3.11 on a PC equipped with an Intel Core i7-12700K  CPU and an Nvidia RTX 3080 GPU. Other system parameters are set as follows: $M=8, L=6$ and $Q=4$  with the number of DFT points $N_{{fft}}=256$, and the number of grid-search points for MUSIC $N_{ {music }}=256$. Together with Figs \ref{fig:changing_P}-\ref{fig:changing_L}, Table \ref{tab:running_time} demonstrates that for the SML estimation, the proposed approach achieves slightly better latency performance than the AE-based method, while for the DML case, the proposed approach significantly outperforms the DFT scheme with the same magnitude of running time, i.e., $10^{-3}$\,s, showing advantageous trade-off between computational complexity and estimation performance.

\section{Conclusions and Remarks}\label{conclusion}
In this letter, we proposed a novel unsupervised training unifying both DML and SML cases for AoD estimation in MISO DL LoS transmissions. In this framework, we proposed to train the deep learning model with the input data including both the received signal and available pilot-sequence information known at the receiver, and to adopt a common neural network architecture that is shared by both SML and DML based AoD estimation. In comparisons with existing benchmarks, numerical results demonstrated that the proposed unsupervised learning approach achieves superior performance in terms of estimation accuracy, while maintaining manageable low estimation overhead. In real-world deployment of the proposed framework, where there may be non-LoS (NLoS) components or out-of-distribution (OOD) input samples involved, fine-tuning the trained model with a much smaller number of practical measurement samples than training samples can be a quick fix for adaption to the new environment. In the future investigation, we may formally incorporate statistical priors into ML estimation formulation to handle highly dynamic NLoS components and leverage advanced techniques such as domain adaption or transfer learning to address OOD issues.

\bibliographystyle{IEEEtran}
\bibliography{ref.bib}

\end{document}